\documentclass[twocolumn,amsmath,amssymb,floatfix,superscriptaddress,prl]{revtex4-2}

\usepackage{bm,graphicx,url,epsf,color}
\usepackage{amssymb}
\usepackage{amsthm}
\usepackage{xcolor}
\usepackage{comment}
\usepackage{relsize}
\usepackage{xspace}
\usepackage[normalem]{ulem}
\usepackage[super]{nth}
\usepackage{braket}
\usepackage{soul}
\usepackage{textcomp}
\usepackage[T1]{fontenc}
\usepackage{stmaryrd}

\graphicspath{{graphics/}}

\usepackage[colorlinks=true,linkcolor=blue,citecolor=blue]{hyperref}

\newcommand{\bd}[1]{{\boldsymbol{#1}}}

\let\origaddcontentsline\addcontentsline

\begin{document}
	
	\title{A discrete realization of the boundary sine-Gordon model and the Schmid transition}
	
	\author{Nicolas Paris}
	\affiliation{Sorbonne Universit\'e, CNRS, Laboratoire de Physique Th\'eorique de la Mati\`ere Condens\'ee, LPTMC, F-75005 Paris, France}
	\affiliation{Universit\'e Paris Cit\'e, CNRS,  Laboratoire  Mat\'eriaux  et  Ph\'enom\`enes  Quantiques, 75013  Paris,  France}
	
	\author{Oscar Bouverot-Dupuis}
	\affiliation{Universit\'{e} Paris Saclay, CNRS, LPTMS, 91405, Orsay, France}
	\affiliation{IPhT, CNRS, CEA, Universit\'{e} Paris Saclay, 91191 Gif-sur-Yvette, France}
	
	\author{Christophe Mora}
	\affiliation{Universit\'e Paris Cit\'e, CNRS,  Laboratoire  Mat\'eriaux  et  Ph\'enom\`enes  Quantiques, 75013  Paris,  France}
	
	\graphicspath{{./figures/}}

	\begin{abstract}
		The phase diagram of the boundary sine-Gordon model (bsG), or equivalently of the resistively shunted Josephson junction, has been the subject of recent debate. We introduce a discrete model that reproduces exactly the properties of the bsG model and can be simulated extremely efficiently using cluster Monte Carlo algorithms. This enables us to explore a broad range of parameters and compute the complete universal crossover between the weak- and strong-tunneling regimes. We establish the self-duality of the model and unambiguously identify the vertical line of critical fixed points characteristic of the Schmid transition, recovering the conformal field theory predictions along this line with high accuracy. We further clarify the essential distinctions between the Schmid and Berezinskii–Kosterlitz–Thouless transitions. Our results definitively establish the existence and nature of the Schmid transition and demonstrate the validity of the Schmid–Bulgadaev scenario.
	\end{abstract}
	
	\maketitle
	
	\paragraph{Introduction ---}
	
	The interplay between quantum fluctuations and dissipation gives rise to some of the simplest and most fundamental examples of quantum phase transitions. A paradigmatic realization is a quantum particle moving in a periodic potential and coupled to an Ohmic bath~\cite{caldeira_1981,caldeira_1983,weiss_1985}. At low dissipation, quantum fluctuations allow the particle to tunnel between neighboring minima and the system remains delocalized, whereas sufficiently strong dissipation suppresses tunneling and localizes the particle. The boundary sine-Gordon (bsG) model provides the canonical field-theoretical description of this problem. It exhibits a quantum phase transition between these two regimes, known as the Schmid transition~\cite{Schmid_1983}. Remarkably, the location of the transition is independent of the depth of the periodic potential~\cite{Bulgadaev_1984}: varying the tunneling amplitude does not shift the critical value of the dissipation. Instead, the transition is characterized by a continuous line of critical fixed points along which the critical properties vary continuously with the strength of the periodic potential.

	An experimentally relevant realization is provided by a resistively shunted Josephson junction (RSJJ), in which a Josephson junction is coupled to an Ohmic resistor~\cite{zwerger_1986,schon_1990,ingold_1994,Grabert_1998,ingold_1999,penttila_2001,houzet_2020,morel_2021,leger_2023,kurilovich_2025,krasin_2026}. In this setting, the competition between quantum phase fluctuations and dissipation gives rise to a transition between a superconducting, low-resistance regime and an insulating, high-resistance regime. In terms of the phase dynamics, the Josephson coupling generates a periodic potential, while the resistor provides the Ohmic dissipative bath. %\np{On peut rajouter l'Hamiltonien RSJJ ici si nécessaire.} 
	The RSJJ therefore realizes the same boundary sine-Gordon physics as the dissipative quantum particle. This correspondence has made the RSJJ a particularly important platform for studying the Schmid transition and its associated transport properties.
	\begin{figure}[t]
		\centering
		\includegraphics[width=\linewidth]{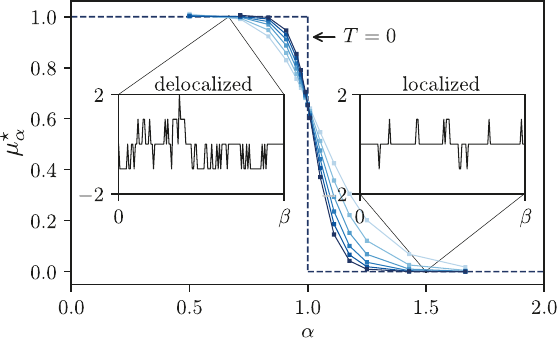}
		\caption{Mobility $\mu_\alpha(2i\pi/\beta)$ as a function of $\alpha$ at $r=1.5$ and $\beta=64,256,1024,4096,16384,65536$ (from light to dark blue). Error bars are smaller than the symbols. When the temperature is decreased, $\mu_\alpha$ converges to $\mu^\star=1$ for $\alpha<1$ and to $\mu^\star=0$ for $\alpha>1$. At $\alpha=1$, it takes an intermediate value that depends on $r$. The dashed line indicates this expected result in the limit $\beta\to\infty$. Typical field configurations of the SOS model \eqref{eq:SOS} at $\beta=128$ are displayed on the left inset for $\alpha=2/3$ and on the right inset for $\alpha=3/2$.
		}
		\label{fig:phase_diagram_quali}
	\end{figure}

	A substantial body of theoretical, numerical, and experimental work has established the Schmid transition and its critical properties. Coulomb-gas analyses~\cite{Bulgadaev_1984}, renormalization-group arguments~\cite{guinea_1985,fisher_1985,Paris_2025}, integrability~\cite{ghoshal_1994,fendley_1994,burshtein_2024}, Monte Carlo simulations~\cite{herrero_2002,kimura_2004,Werner_2005,Lukyanov_2007}, and exact diagonalization~\cite{giacomelli_2024} support the scenario in which the superconducting and insulating phases are separated by a vertical line of critical fixed points with continuously varying critical properties. Remarkably, conformal field theory (CFT) provides an exact description of this critical line~\cite{callan_1994_prl,callan_1994,polchinski_1994,hasselfield_2006,Paris_2025}. Experimental signatures of the transition have also been reported in resistively shunted Josephson junctions~\cite{kuzmin_2025,subero_2026}.
	
	This established picture has nevertheless been questioned by experiments that did not observe the expected insulating behavior~\cite{yagi_1997,penttila_1999,subero_2023,murani_2020,murani_comment,murani_reply} and by theoretical studies reporting deviations from the conventional transition line~\cite{masuki_2022,daviet_2023}. These apparent discrepancies were traced to finite-temperature and finite-frequency effects, which have been shown to account for the experimental observations within the Schmid scenario~\cite{houzet_2024,yamamoto_2024,yeyati_2024}. Likewise, deviations obtained within functional renormalization-group approaches may arise from the approximations involved in the truncation scheme~\cite{daviet_2023}, while the corresponding numerical renormalization group results have been challenged on methodological grounds~\cite{masuki_comment,masuki_reply}. More recently, a Monte Carlo study has confirmed the existence of a transition but interpreted it as belonging to the Berezinskii--Kosterlitz--Thouless (BKT) universality class~\cite{Capone_2026}, a point that we address in the present manuscript.

	In this work, we introduce a discrete model that reproduces the universal behavior of the boundary sine-Gordon model. We simulate it using a highly efficient cluster Monte Carlo algorithm~\cite{Swendsen_1987,Wolff_cluster_algo} adapted to long-range interactions~\cite{Fukui_Todo_2009,ClockFMetManonMichel}, which avoids critical slowing down and enables simulations at unprecedented large system sizes across a broad range of parameters. This allows us to go beyond previous numerical studies~\cite{Werner_2005} and compute universal mobility crossover functions across the weak- and strong-tunneling regimes. These functions enable us to identify unambiguously the localized and delocalized phases and, remarkably, to demonstrate numerically the self-duality of the bsG model. We further identify the vertical line of critical fixed points characteristic of the Schmid transition and recover the exact CFT results along this line with high accuracy. Our results establish the existence and nature of the Schmid transition and confirm the Schmid--Bulgadaev scenario.

	\paragraph{Model and phase diagram ---}
	
	The boundary sine-Gordon model is defined by 
	\begin{align}\label{eq:bsg}
		S_{\rm bsG}=&\frac{\alpha}{8\pi^2 }\underset{|\tau-\tau'|>\tau_c}{\int_0^\beta \!\!d\tau d\tau'}\frac{(\phi(\tau)-\phi(\tau'))^2}{d(\tau-\tau')^2}-\tilde r\int_0^\beta \!\! d\tau \cos\phi(\tau),
	\end{align}
	where $d(\tau)=\beta/\pi \sin(\pi \tau/\beta )$ is the chord length, $\tau, \tau'$ are imaginary times, $\beta$ is the inverse temperature and $\tau_c$ an ultraviolet cut-off. It describes the physics of a dissipative particle of coordinate $\phi$ moving in a periodic potential with parameter $\tilde r$. The correspondence with the RSJJ is detailed in the Supplemental Material~\cite{refSM}.
	
	In the large-$\tilde r$ limit, $\phi$ becomes localized at the minima of the cosine potential, $\phi_{\rm min}=2\pi n$, with $n\in\mathbb{Z}$, and can tunnel from one minimum to another via instantons. Neglecting small fluctuations around these minima and discretizing imaginary time as $\tau\to i\tau_c$ leads to the solid-on-solid (SOS) model (with $\tau_c=1$ from now on)
	\begin{align}\label{eq:SOS}
		S_{\rm SOS}=\frac{1}{2}\sum_{\substack{i,j=1\\ i \ne j}}^\beta \mathcal{D}_r(i-j) (n_i-n_j)^2,
	\end{align}
	where the inverse temperature $\beta$ is expressed in units of $\tau_c$ and is therefore a dimensionless integer. At each site $i$, the field $n_i$ takes integer values that label the minima of the cosine potential. The kernel $\mathcal{D}_r(i)$ must obey the asymptotic form
	\begin{equation}
		\mathcal{D}_r(i) \overset{i \gg 1}{=} \frac{\alpha}{d(i)^2}
	\end{equation}
	to reproduce the long-range interaction of the bsG. The short-time behavior of $\mathcal{D}_r(i)$ does not affect the universal properties but can be tuned to favor or penalize the creation of instantons, much like the parameter $\tilde r$ in Eq.~\eqref{eq:bsg}. To reproduce such a behaviour, we explicitly consider the ansatz
	\begin{equation}\label{eq:kernel}
		\mathcal{D}_r(i)=\frac{\alpha(1+(r-1)\delta_{|i|,1})(1-\exp(-r d(i)))}{ d(i)^2}.
	\end{equation}
	Increasing (decreasing) $r$ amounts to strengthening (weakening) the coupling in front of $(n_i - n_j)^2$ in Eq.~\eqref{eq:SOS} at small separation. This penalizes (favors) instantons which are kinks in the field $n$ (see~\cite{refSM}).
	
	We expect the SOS model~\eqref{eq:SOS} and the bsG model~\eqref{eq:bsg} to belong to the same universality class, as they share the same symmetries, dimensionality and long-range interaction. Moreover, various SOS models have been successfully used to study field theories similar to the bsG model~\cite{Chui_1976,Hasenbusch_1994,Schehr_2005,paris2026}. The main advantage of the SOS model over the field theory is that it can be simulated more efficiently using cluster Monte Carlo algorithms adapted to long-range interactions, as in~\cite{paris2026}.
	
	The transport properties of the bsG model are directly related to the mobility~\cite{Schmid_1983}, which quantifies the degree of localization of the field $\phi$:
	\begin{equation}\label{eq:mob_bsg}
		\mu_\alpha(i\omega_m)=\frac{\alpha|\omega_m|}{2\pi}\langle\phi(i\omega_m)\phi(-i\omega_m)\rangle,
	\end{equation}
	with $\omega_m = 2\pi m /\beta$ a Matsubara frequency. Replacing $\phi \to 2\pi n$, it can be equivalently written as 
	\begin{align}
		\mu_\alpha(i\omega_m)=2\pi \alpha|\omega_m|\langle n(i\omega_m)n(-i\omega_m)\rangle
	\end{align}
	in the SOS model. As shown in Fig.~\ref{fig:phase_diagram_quali}, its zero-frequency value, denoted by $\mu^\star_\alpha=\mu_\alpha(0)$, is an order parameter for the Schmid transition. It distinguishes between the delocalized phase ($\mu^\star=1$) at $\alpha<1$ where the particle can tunnel from one minimum of the potential to another and the localized phase ($\mu^\star=0$) at $\alpha>1$ where the particle is stuck in a single minimum. In the context of the RSJJ, the phase at $\alpha<1$ is called insulating due to the lack of current flowing through the Josephson junction (JJ), while for $\alpha>1$ the JJ is superconducting. 
	
	The Schmid transition is peculiar for several reasons. First, it exhibits a duality between $\alpha$ and $1/\alpha$ which we review and check numerically. Second, the critical point at $\alpha=1$ is actually a vertical line of fixed points~\cite{guinea_1985} with intermediate mobilities $\mu^\star\in [0,1]$.

	\paragraph{Duality $\alpha \leftrightarrow 1/\alpha$ ---}
	
\begin{figure}
		\centering
		\includegraphics[width=0.495\linewidth]{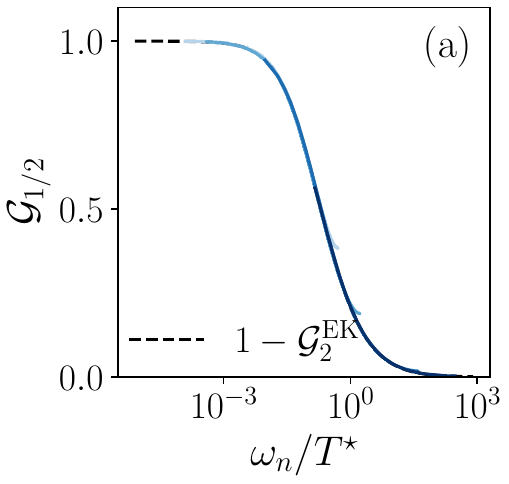}
		\includegraphics[width=0.482\linewidth]{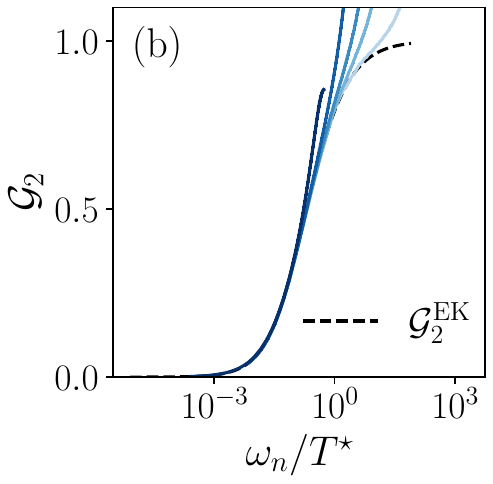}\\
		
		\includegraphics[width=0.49\linewidth]{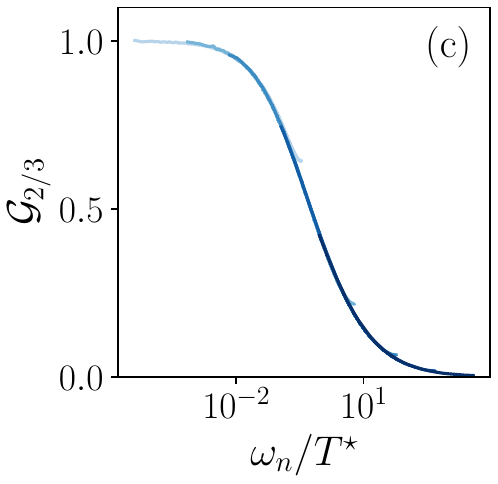}
		\includegraphics[width=0.495\linewidth]{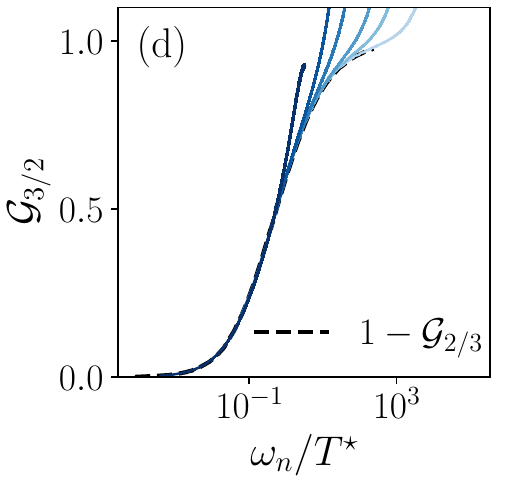}
		
		\caption{(a): Mobility of the SOS model for $\alpha=1/2$ and for various values of $r$ in terms of the rescaled frequency. The dashed black line corresponds to $1-\mathcal{G}^{\rm EK}_2$. (b): Mobility of the SOS model for $\alpha=2$. The black dashed line corresponds to the Emery-Kivelson solution~\eqref{eq:EK}. The deviations from the exact result at high-frequency ($\omega_n \simeq 1$) are non-universal effects. (c): Mobility of the SOS model for $\alpha=2/3$. (d): Mobility of the SOS model for $\alpha=3/2$. The black dashed line corresponds to $1-\mathcal{G}_{2/3}$ where $\mathcal{G}_{2/3}$ is the universal mobility crossover at $\alpha=2/3$ extracted from (c).}
		\label{fig:EK}
	\end{figure}

	As first argued by Schmid and Bulgadaev \cite{Schmid_1983,Bulgadaev_1984}, the coupling $\tilde r$ in Eq.~\eqref{eq:bsg} is irrelevant in the renormalization group sense for $\alpha<1$. For $\alpha>1$, $\tilde r$ flows to strong coupling and model \eqref{eq:bsg} becomes perturbatively untractable. It was however proposed by Schmid, and later supported by circuit quantum electrodynamics~\cite{ingold_1999,Paris_2025,Giacomelli_2026} and integrability arguments~\cite{Fendley_1995,fendley_1998,fendley_1999}, that the bsG model is self-dual upon exchanging $\alpha$ and $1/\alpha$, so that the physics at $\alpha<1$ can be recovered from that at $\alpha>1$. This can be seen in the mobility which is expected to exhibit a universal scaling behavior at $\alpha\ne 1$ as a function of the rescaled frequency $\omega/T^\star$, i.e. 
	\begin{equation}
		\mu_\alpha(i\omega)\underset{|\omega|\ll 1}{=}\mathcal G_\alpha\left(\frac{\omega}{T^\star}\right)
	\end{equation}
	for $|\omega|\ll 1$ (recall that the UV cutoff is set to 1), where $T^\star$ is a $r$-dependent crossover scale and $\mathcal G_\alpha$ a universal function. The duality condition then reads
	\begin{equation}\label{eq:duality}
		\mathcal G_\alpha\left(\frac{\omega}{T^\star}\right)=1-\mathcal{G}_{1/\alpha}\left(\frac{\omega}{T^\star}\right).
	\end{equation}
	To test this prediction, we first concentrate on the model at $\alpha=2$ and $\alpha=1/2$ since these cases are exactly solvable. Indeed, the bsG being equivalent to the two-channel Kondo model, Emery-Kivelson's refermionization \cite{Emery_Kivelson_1992,sela2011,mitchell2012} (see also the Sup. Mat. of \cite{paris2026}) provides the exact result at $\alpha=2$
	\begin{equation}\label{eq:EK}
		\mathcal G^{\rm EK}_{2}\left(\frac{\omega}{T^\star}\right)=1-\frac{T^\star\ln\left(1+| \omega|/T^\star\right)}{| \omega|},
	\end{equation}
	with $T^\star\propto e^{- r}$. Moreover, refermionization at $\alpha=1/2$ shows that the universal mobility crossover is exactly given by $\mathcal{G}_{1/2}=1-\mathcal{G}^{\rm EK}_2$ (see~\cite{refSM}). The universal scaling functions obtained from Monte Carlo simulations of the SOS model are in complete quantitative agreement with these exact results (see Fig.~\ref{fig:EK}).
	
	Moving away from the previous exactly solvable point, we have also checked the duality condition~\eqref{eq:duality} for $\alpha=2/3$ and $\alpha=3/2$ in Fig.~\ref{fig:EK}. Our Monte Carlo approach therefore conclusively establishes the $\alpha\leftrightarrow 1/\alpha$ duality for generic values of $\alpha$.

	\paragraph{Vertical line of fixed point ---}
	
	The line of fixed points at $\alpha=1$ is described by a single parameter continuous family of CFTs parametrized by $\mu^\star\in[0,1]$~\cite{callan_1994_prl,callan_1994,polchinski_1994,Lukyanov_2007,Paris_2025}. CFT approaches give access to the spectrum of the theory~\cite{hasselfield_2006}, and consequently to all thermodynamic observables, but also to certain correlation functions, such as 
	\begin{align}
		G_q(\tau)=\langle e^{iq\phi(\tau)}e^{-iq\phi(0)}\rangle
	\end{align}
	for $q\in [0,1/2[$~\cite{Werner_2005,Lukyanov_2007}. At large time, its universal behavior is captured by
	\begin{align}\label{eq:eta_asymptotic}
		G_q(\tau)\propto d(\tau)^{-2\eta_q}.
	\end{align}
	The decay exponent $\eta_q$ is known to be equal to~\cite{Lukyanov_2007}
	\begin{equation}\label{eq:eta_q}
		\eta_q=\frac{1}{\pi^2}\arcsin^2\left(\sqrt{\mu^\star}\sin(\pi q)\right),
	\end{equation}
	where $\mu^\star$ is the zero-temperature mobility. 
	
\begin{figure}
		\centering
		\includegraphics[width=0.49\linewidth]{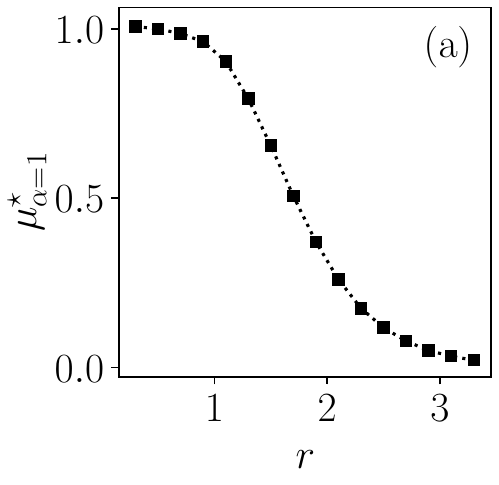}
		\includegraphics[width=0.49\linewidth]{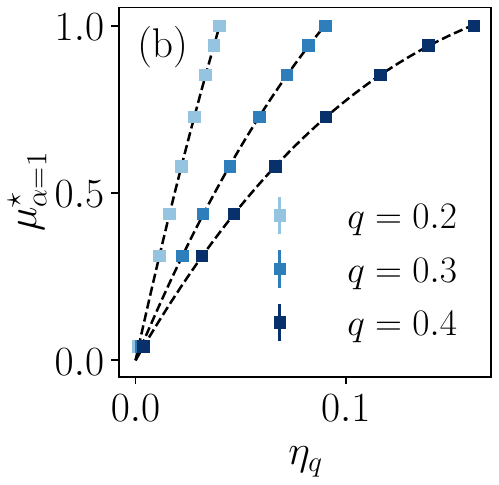}
		\caption{(a) Mobility $\mu^\star_{\alpha=1}$ as a function of $r$ at $\beta=8192$. The dotted line is a guide to the eye. (b) Mobility $\mu^\star_{\alpha=1}$  as a function of $\eta_q$ for $q=0.2,0.3,0.4$. The Monte Carlo data (blue squares) was obtained for $\beta=8192$. The dashed lines indicate the exact result~\eqref{eq:eta_q}. Error bars are smaller than the symbols. The agreement is excellent despite the finite temperature.}
		\label{fig:mu_eta_q}
	\end{figure}

	Within our approach, we observe that varying continuously $r$ at $\alpha=1$ gives access to the vertical line of fixed points with intermediate mobility values interpolating between $0$ and $1$ (see Fig.~\ref{fig:mu_eta_q}(a)). Due to the broad range of values of $r$ accessible, we can obtain the complete curve of $\mu^\star$ ranging from $0$ to $1$ as a function of $r$. Yet, since $r$ is a microscopic model-dependent parameter, we prefer plotting $\mu^\star$ as a function of $\eta_q$ to obtain a universal curve (see Fig.~\ref{fig:mu_eta_q}(b)). The agreement with the exact result~\eqref{eq:eta_q} is excellent.
	We emphasize that this definitely concludes about the existence and nature of the Schmid transition in the boundary sine-Gordon model.

	\paragraph{Similarities and differences between Schmid and BKT transitions ---}
	
	The Schmid transition shares common features with the BKT transition. Indeed, a small $\tilde r$-expansion maps the bsG onto a 1D Coulomb gas of logarithmically-interacting $\pm 1$ charges, which undergoes a binding-unbinding transition at the Schmid transition. Such a transition is commonly associated with BKT physics and also occurs in the sine-Gordon model whose Coulomb gas is the 2D analogue of the 1D bsG one, or in the Kondo model (equivalently the one-dimensional long-range Ising model \cite{anderson1969}) which also maps onto a 1D gas (see~\cite{refSM} for derivations). The key distinction between the bsG and sine-Gordon models lies in the dimensionality of their Coulomb gas representations, whereas the distinction between the bsG and Kondo models lies in the ordering of the charges. In the Kondo model, positive and negative charges must alternate, while no such constraint applies in the bsG model. It is the combination of dimensionality and the (lack of) charge ordering which defines the Schmid universality.
	
\begin{figure}
		\centering
		\includegraphics[width=\linewidth]{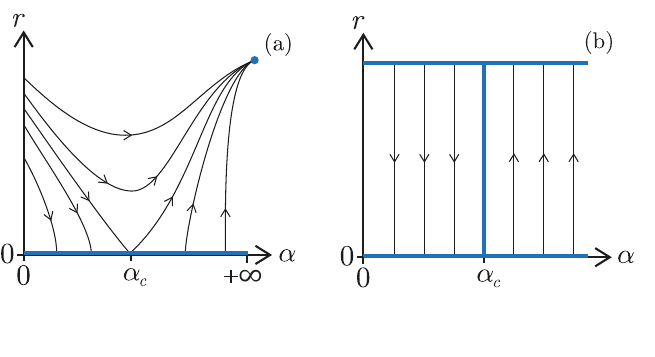}
		\caption{(a): Flow diagram of the BKT transition corresponding to Eqs.~(\ref{eq:BKT1},\ref{eq:BKT2}). (b): Flow diagram of the Schmid transition corresponding to Eqs.~(\ref{eq:Schmid1},\ref{eq:Schmid2}). Fixed points are in blue.}
		\label{fig:BKT_vs_schmid}
	\end{figure}

	From a renormalization-group (RG) point-of-view, the two transitions also exhibit different structures. For the BKT transition, the Coulomb-gas representation leads to two coupled running parameters, the defect fugacity $r$ and the effective stiffness $\alpha$. To lowest nontrivial order, their RG equations take the form
	\begin{align}
		\frac{dr}{dl} &= (\alpha-\alpha_c)r, \label{eq:BKT1}\\
		\frac{d\alpha}{dl} &= cr^2, \label{eq:BKT2}
	\end{align}
	where $c>0$ is model-dependent. The line of fixed points at $r=0$ and $\alpha<\alpha_c$ is attractive and stable under the RG flow. The BKT transition occurs at the endpoint $\alpha=\alpha_c$ of this line and for $\alpha>\alpha_c$, the flow goes to a strong-coupling fixed point at $\alpha=+\infty$ (see Fig.~\ref{fig:BKT_vs_schmid} (a)). At the transition, the renormalized stiffness therefore takes the universal value $\alpha_c$. This gives rise, for instance, to the universal Nelson--Kosterlitz jump of the helicity modulus in the two-dimensional XY model.
	
	The Schmid transition is superficially similar. In the Coulomb-gas representation of the bsG model~\eqref{eq:bsg}, the charge fugacity $\tilde r$ also goes from relevant to irrelevant at a critical coupling $\alpha_c$. Perturbatively close to the transition, one finds \cite{Bulgadaev_1984,Paris_2025}
	\begin{align}
		\frac{d\tilde r}{dl} &= \tilde r\left(\frac{1}{\alpha_c}-\frac{1}{\alpha}\right), \label{eq:Schmid1}\\
		\frac{d\alpha}{dl} &= 0. \label{eq:Schmid2}
	\end{align}
	The second equation is valid to all orders in perturbation and is the essential difference. In the bsG model, the coefficient $\alpha$ of the long-range interaction in the Gaussian action is not renormalized by the boundary cosine, in contrast to the BKT flow. The perturbative RG equations therefore predict a vertical transition line at $\alpha=\alpha_c$: for $\alpha<\alpha_c$ the cosine is irrelevant and the system flows towards the delocalized fixed-point line $\tilde r=0$, whereas for $\alpha>\alpha_c$ it flows towards strong coupling (see Fig.~\ref{fig:BKT_vs_schmid} (b)).
	
	As explained above, the transition line remains at $\alpha=\alpha_c$ even beyond the perturbative regime, and is a family of fixed points which continuously depend on the microscopic value $\tilde r$. Thus, right at the Schmid transition, the behavior of observables (e.g. the algebraic decay of two-point correlation functions) depends explicitly on $\tilde r$, while for the BKT scenario there is no such remaining freedom. This distinction is summarized schematically in Fig.~\ref{fig:BKT_vs_schmid}. In the BKT case, the transition is associated with the endpoint of a line of Gaussian fixed points, and the renormalized couplings at this endpoint are unique. In the Schmid case, the transition itself forms a line of fixed points. The two transitions thus share the same logarithmic Coulomb-gas interaction and the proliferation of topological defects, but differ in the structure of their infrared fixed points and, consequently, in the universality of their critical observables.
	
	\paragraph{Conclusion ---}
	We have introduced a discrete model related to the boundary sine-Gordon model and shown that it belongs to the same universality class. The model can be sampled highly efficiently using cluster Monte Carlo algorithms, completely eliminating critical slowing down. This approach provides direct access to the full universal mobility crossover over a broad range of parameters. In particular, our results demonstrate the duality between the weak- and strong-tunneling regimes and provide compelling evidence for the vertical line of critical fixed points in the phase diagram, which is a defining feature of the Schmid transition. We further clarify the distinction between the Schmid and BKT transitions, despite their common features.

	\paragraph{Acknowledgements ---}We acknowledge fruitful discussions with N. Dupuis. C.~M. acknowledges funding from the Agence Nationale de la Recherche under the France 2030 programme, reference ANR-22-PETQ-00122 (EQUBITFLY). O.B.-D. acknowledges the support of the French ANR under the grant ANR-22-CMAS-0001 (\emph{QuanTEdu-France} project).

\renewcommand{\addcontentsline}[3]{}

\let\addcontentsline\origaddcontentsline  % Restore

\clearpage
\newpage 

\onecolumngrid

\begingroup  % <---- Start a confined formatting scope

\setcounter{equation}{0}
\setcounter{figure}{0}
\setcounter{table}{0}
\setcounter{page}{1}
\renewcommand{\theequation}{S\arabic{equation}}
\renewcommand{\theHequation}{S\arabic{equation}}
\renewcommand{\thefigure}{S\arabic{figure}}	
\renewcommand{\theHfigure}{S\arabic{figure}}
\renewcommand{\thetable}{S\arabic{table}}	
\renewcommand{\theHsection}{S\arabic{section}}
\renewcommand{\theHsubsection}{S\arabic{section}.\Alph{subsection}}
\renewcommand{\theHsubsubsection}{S\arabic{section}.\Alph{subsection}.\arabic{subsubsection}}

% Citation numbering override just for SM
\renewcommand{\bibnumfmt}[1]{[S#1]}
\renewcommand{\citenumfont}[1]{S#1}

% Title
\centerline{\bf Supplemental Material: A discrete realization of the boundary sine-Gordon model and the Schmid transition}
\medskip
\centerline{Nicolas Paris, Oscar Bouverot-Dupuis, and Christophe Mora}

\bigskip 

In this Supplemental Material, we detail the correspondence between the parameters of the boundary sine-Gordon (bsG) model and those of the resistively shunted Josepshon junction (RSJJ). We then explain our choice regarding the kernel of the SOS action and derive explicitly the mobility at $K=1/2$ via refermionization. Finally, we review the Coulomb gas expansions of models belonging either to the Schmid or to the BKT universality class.

\tableofcontents

\section{\label{app:RSJJ}I. Relation with the RSJJ}
The RSJJ model consists of a Josephson junction with Josephson energy $E_J$ and charging energy $E_C=e^2/(2C)$, shunted by a resistance $R$. The resistive environment can be modeled as a transmission line of length $L$, propagation velocity $v$, and characteristic impedance $R$. The Hamiltonian of the circuit is
\begin{align}\label{eq:HRSJJ}
	\hat H=4E_C(\hat N-\hat n)^2-E_J\cos\hat\varphi+\sum_{q\geq0}\hbar \omega_q\hat a_q^\dagger\hat a_q.
\end{align}
Here, $\hat N$ is the charge transferred across the junction (in units of $2e$) and $\hat\varphi$ is the canonically conjugate phase difference. The operators $\hat a_q^\dagger$ and $\hat a_q$ create and annihilate photonic excitations in the transmission line with mode frequency $\omega_q=(q+1/2)\Delta$ with $\Delta=\pi v/L$ the mean level spacing of the transmission line. The operator
\begin{align}
	\hat n=\frac{1}{\pi}\sum_{q\geq0}\sqrt{\frac{R_Q \Delta}{2R\omega_q}}(\hat a_q^\dagger+\hat a_q)
\end{align}
is the charge displaced from the transmission line to the junction (in units of $2e$), with $R_Q=h/4e^2$ the resistance quantum.

In the limit of a semi-infinite transmission line, $L\to\infty$, the level spacing vanishes and the environmental modes form a continuum. Integrating out these modes yields the action
\begin{align}\label{eq:bsg2}
	S=&\frac{\alpha}{8\pi^2 }\int_0^\beta d\tau d\tau'\frac{(\phi(\tau)-\phi(\tau'))^2}{d(\tau-\tau')^2} + \int_0^\beta d\tau \left[\frac{1}{16E_C}(\partial_\tau\phi)^2-E_J\cos\phi(\tau)\right],
\end{align}
with $\alpha=R_Q/R$.
The ultraviolet cutoff implementation, either via a hard frequency cutoff, as in Eq.~(1) of the main text or via the charging-energy term proportional to $(\partial_\tau\phi)^2$, as in Eq.~\eqref{eq:bsg2}, does not affect the universal properties of the system. The correspondence between the parameters of the RSJJ and those of the bsG action is summarized in Table.~\ref{tab:corr}. The validity of the bsG description for the RSJJ has been established by comparing exact diagonalization of the finite-length Hamiltonian~\eqref{eq:HRSJJ} with CFT predictions for the bsG model~\cite{giacomelli_2024_sm,Paris_2025_sm}.

\begin{table}
	\caption{Correspondence between parameters of the RSJJ and bsG models.}
	\label{tab:corr}
	\begin{ruledtabular}
		\begin{tabular}{cccc}
			&RSJJ & bsG& \\
			\colrule
			&$E_J$ & $\tilde r$ &\\
			&$E_C$ & $\tau_c^{-1}$& \\
			&$R_Q/R$ & $\alpha$& \\
		\end{tabular}
	\end{ruledtabular}
\end{table}

\section{\label{app:kernel}II. Kernel of the SOS model and universality}
The kernel of the SOS model (Eq.~(4) of the main text) is plotted in Fig.~\ref{fig:kernel} for various values of $r$. As displayed in the figure, decreasing $r$ favors fluctuations of the field $n$, i.e. instantons. Exactly at $r=0$, the kernel vanishes and the field is completely free. When $r\to\infty$, $\mathcal{D}_r(1)$ diverges and the field $n$ becomes flat. 
\begin{figure}[h!]
	\centering
	\includegraphics[width=0.5 \linewidth]{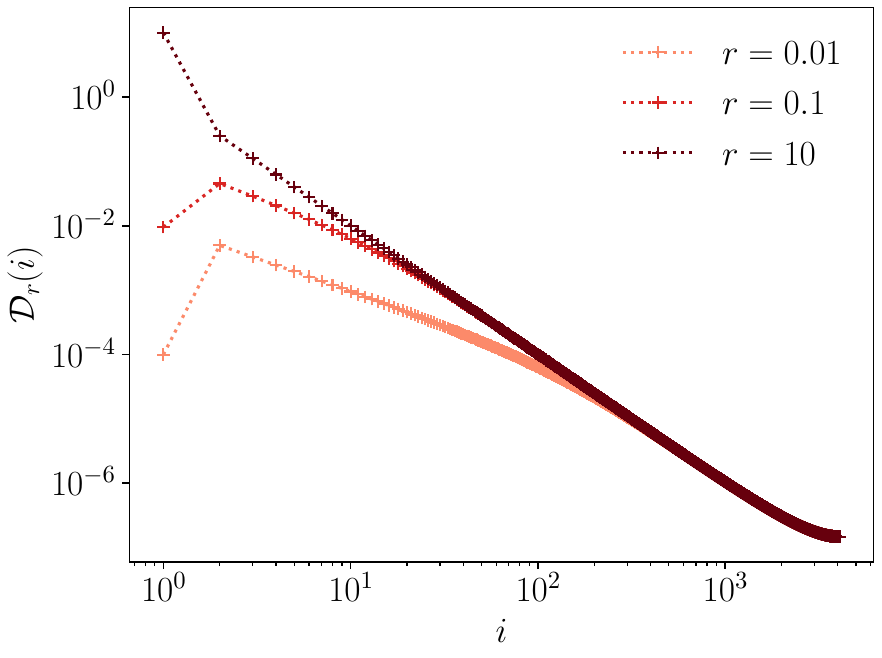}
	\caption{Kernel $\mathcal D_r$ as a function of the distance at $\beta=8192$, $\alpha=1$ and for various values of $r$. The long distance behavior is identical whatever the value of $r>0$.}
	\label{fig:kernel}
\end{figure}

The more natural kernel choice 
\begin{align}
	\mathcal{D}(i)=\frac{\alpha(1+(r-1)\delta_{|i|,1})}{d(i)^2},
\end{align}
was implemented in Ref.~\cite{paris2026_sm} in the case of the multichannel Kondo model. However, it is numerically less convenient as the fluctuations of the field are still strongly penalized at $r=0$.

\section{III. Refermionization at $\alpha=1/2$}
\label{app:refermionization}

This section shows that the mobility $\mu_\alpha(i\omega)$ at $\alpha=1/2$ is given by $1-\mathcal G^{\rm EK}_{2}(\omega/T^\star )$ with $\mathcal G^{\rm EK}_2$ given in Eq.~(9) of the main text.

\subsection{A. Hamiltonian formulation}
After rescaling $\phi \to 2\phi$, the bsG model at $\alpha=1/2$ reads
\begin{align}
	S_{\rm bsG}=&\frac{1}{4\pi^2} \int d\tau d\tau'\frac{(\phi(\tau)-\phi(\tau'))^2}{d(\tau-\tau')^2}-\tilde r\int d\tau \cos2\phi(\tau).
\end{align}
It can be recovered from the $(1+1)$d chiral boson $\phi(x,\tau)$ action
\begin{equation}
	S = \int dx d\tau \left\{\frac{1}{4\pi} \left[(\partial_x \phi)^2 + i \partial_x \phi \partial_\tau \phi\right] - \delta(x) \tilde r \cos(2\phi)\right\}
\end{equation}
by integrating out the degrees of freedom at $x\ne 0$ \cite{Gogolin_2004_sm}. Going back to the Hamiltonian formalism gives
\begin{equation}\label{eq:H_K2}
	H = \frac{1}{4\pi} \int dx \left\{(\partial_x \phi)^2 - \delta(x) \tilde r \cos(2\phi)\right\},
\end{equation}
where $\phi$ is a chiral boson such that $[\phi(x),\phi(x')]=i\pi {\rm sign}(x-x')$.

\subsection{B. A bosonization identity}
To refermionize the Hamiltonian \eqref{eq:H_K2}, we use the identity
\begin{equation}\label{eq:refermionization_K2}
	\psi(0)\partial_x \psi(0) + \psi^\dagger(0) \partial_x \psi^\dagger(0)=\frac{i}{\pi a^2}\cos(2\phi(0)),
\end{equation}
which we prove in the following.

We start from the bosonization identity \cite{vonDelft_1998_sm}
\begin{align}\label{eq:bosonization_identity}
	\psi(x)=&\frac{\eta}{\sqrt{L}} :e^{- i\phi(x)}:=\frac{\eta}{\sqrt{2\pi a}} e^{ - i\phi(x)},
\end{align}
where the field $\phi(x)=\phi^0 - 2\pi N \frac{x}{L}  + \varphi^\dagger(x) +\varphi(x)$ includes the zero mode $\phi^0$ which, if we neglect all $1/L$ contributions, can be assumed to commute with the particle number $N$. $\eta$ is a Majorana fermion such that $\eta^2=1$, $a$ is a UV cutoff and $:\bullet:$ denotes normal ordering where all $\varphi^\dagger$ are put to the left of all $\varphi$. We wish to compute
\begin{equation}
	\psi(0) \partial_x \psi(0)=\lim_{\varepsilon \to 0} \psi(0)\frac{\psi(\varepsilon)-\psi(0)}{\varepsilon}=\lim_{\varepsilon \to 0} \frac{\psi(0)\psi(\varepsilon)}{\varepsilon},
\end{equation}
since $\psi(0)^2=0$ due to Fermi statistics. We are therefore left with the standard task of evaluating the point-split product
\begin{align}
	\psi(0)\psi(\varepsilon)=&\frac{1}{L} :e^{- i\phi(0)}: :e^{- i\phi(\varepsilon)}:\\
	=&\frac{1}{L}e^{-i2\phi^0 +i 2\pi N \frac{\varepsilon}{L} }e^{-i\varphi^\dagger(0)}e^{-i\varphi(0)}e^{-i\varphi^\dagger(\varepsilon)}e^{-i\varphi(\varepsilon)},\nonumber
\end{align}
where we have used $\eta^2=1$. To make the operators meet at $\varepsilon=0$ one has to remove the UV divergences by normal ordering everything. This is done using the formula
\begin{equation}\label{eq:normal_order}
	e^{-i\varphi(0)}e^{-i\varphi^\dagger(\varepsilon)}=e^{-i\varphi^\dagger(\varepsilon)}e^{-i\varphi(0)}e^{-[\varphi(0),\varphi^\dagger(\varepsilon)]}.
\end{equation}
Combining this with $[\varphi(x),\varphi^\dagger(x')] = - \ln[-i\frac{2\pi}{L}\big(x-x' + i a \big)]$ (see \cite{vonDelft_1998_sm}) yields
\begin{align}
	\psi(0)\psi(\varepsilon)=&\frac{i2\pi \varepsilon}{L^2} :e^{- i\phi(0)} e^{- i\phi(\varepsilon)}: = \frac{i2\pi \varepsilon}{L^2} :e^{- i2\phi(0)}: + O(\varepsilon^2).
\end{align}
Finally, taking $\varepsilon \to 0$ and removing the normal ordering of the exponential in a similar way to Eq.~\eqref{eq:normal_order} yields
\begin{align}
	\psi(0)\partial_x \psi(0) =&\frac{i2\pi}{L^2} :e^{- i2\phi(0)}: = \frac{i}{2\pi a^2} e^{- i2\phi(0)}.
\end{align}
Adding the hermitian conjugate leads to Eq.~\eqref{eq:refermionization_K2}.

\subsection{C. Diagonalization of the fermionized Hamiltonian}
Using Eq.~\eqref{eq:refermionization_K2}, the Hamiltonian \eqref{eq:H_K2} can be refermionized into two Majorana fermions 
\begin{equation}
	H = \frac{i}{2} \int dx \left\{\psi^\dagger\partial_x \psi + i \delta(x) \lambda \left(\psi \partial_x \psi + \psi^\dagger \partial_x \psi^\dagger\right)\right\},
\end{equation}
with $\psi$ a chiral fermion and $\lambda=\pi a^2 \tilde r$.

Going to the path integral formalism and introducing Majorana fermions such that $\psi=\frac{1}{\sqrt{2}}(\xi+i\eta)$ the action reads $S=S[\xi,\lambda] + S[\eta,-\lambda]$ with
\begin{align}
	S[\xi,\lambda]=&\frac12 \int_{\omega,k,k'} \xi(i\omega,k)\bigg[2\pi\left(i\omega + k\right)\delta(k-k') + \lambda (k+k')\bigg]\xi(-i\omega,-k'),
\end{align}
where $\xi(\tau,x) = \int_{k, \omega} e^{-i\omega \tau+ikx}\xi(i\omega,k)$ and $\int_k = \int \frac{dk}{2\pi}$, $\int_\omega=\int \frac{d\omega}{2\pi}$. The inverse propagator is a diagonal matrix with a rank-2 perturbation in momentum space so it can be inverted using Woodbury's identity. This yields
\begin{equation}
	G^\xi(i\omega,k,k') = \frac{2\pi \delta(k-k')}{d_k} + \lambda \frac{I_0 k k' - (I_1 + 1)(k+k') + I_2}{[ (I_1 + 1)^2 - I_0 I_2] d_k d_{k'}} = -\langle \xi(i\omega,k) \xi(-i\omega,-k') \rangle,
\end{equation}
with $I_n = \lambda \int_{k\in [-\Lambda,\Lambda]} k^n/d_k$ and $d_k=i \omega + k$. Defining $G^\xi_{m n}(i\omega) = \int_{k,k'} G^\xi(i\omega,k,k') k^m {k'}^n$, one finds
\begin{align}
	\begin{pmatrix}
		G^\xi_{00}&G^\xi_{01}\\
		G^\xi_{10}&G^\xi_{11}
	\end{pmatrix} &= \frac{1}{\lambda N^\xi}\begin{pmatrix}
		I_0 & (I_1 + 1)I_1 - I_0 I_2\\
		(I_1 + 1)I_1 - I_0 I_2 & I_2
	\end{pmatrix} =  i \int d \tau e^{i\omega \tau} \begin{pmatrix}
		i\langle \xi(\tau)\xi(0)\rangle & -\langle \xi(\tau)\partial_x\xi(0)\rangle\\
		\langle \partial_x \xi(\tau) \xi(0)\rangle & i\langle \partial_x\xi(\tau)\partial_x\xi(0)\rangle
	\end{pmatrix}
\end{align}
where all fields are evaluated at $x=0$ and
\begin{align}
	&I_0 = \frac{\kappa}{i\omega} f(\omega), \quad I_1=\kappa \left(1-f(\omega)\right), \quad I_2= -i\omega I_1,\\
	&N^\xi=(I_1 + 1)^2 - I_0 I_2=(1+\kappa)^2 +(1-(1+\kappa)^2)f(\omega),
\end{align}
with $f(\omega) = \omega/\Lambda\arctan(\Lambda/\omega)$ and $\kappa=\lambda\Lambda/\pi$. All quantities for $\eta$ are obtained by substituting $\lambda,\kappa \to - \lambda, -\kappa$.

As shown in the Sup. Mat. of Ref.~\cite{paris2026_sm}, the mobility $\mu(i\omega)=\frac{|\omega|}{\pi }\langle\phi(i\omega)\phi(-i\omega)\rangle$ of the bsG can be written as
\begin{align}
	\mu(i\omega)&=1-\frac{4\pi r}{|\omega|}\langle\cos(2\phi(0))\rangle+\frac{4\pi r^2}{|\omega|}\int_0^\beta d\tau\, e^{i\omega \tau}\langle \sin(2\phi(\tau))\sin(2\phi(0))\rangle,
\end{align}
which, in the fermionic variables, becomes
\begin{align}
	\mu(i\omega)=&1 -\frac{4\pi \lambda}{|\omega|}\int_{\omega'} \left[G_{01}^\xi(i\omega')-G_{01}^\eta(i\omega')\right] + \frac{4\pi \lambda^2}{|\omega|}\int_{\omega'} \left[G_{00}^\eta(i\omega') G_{11}^\xi(i\omega-i\omega') - G_{10}^\eta(i\omega') G_{01}^\xi(i\omega-i\omega') + (\xi \leftrightarrow \eta) \right]\nonumber\\ \label{eq:mu_full_expr}
	=&1  - \frac{4\kappa^2(1-\kappa^2)}{|\omega|}\int d\omega'\,\left\{\frac{(1-f(\omega'))(1-f(\omega-\omega'))}{N_\eta(\omega')N_\xi(\omega-\omega')} - \frac{(1-f(\omega'))^2}{N_\eta(\omega')N_\xi(\omega')}\right\}\nonumber\\
	&-\frac{2\kappa^2}{|\omega|}\int d\omega'\left\{\left[\frac{1}{N_\xi(\omega')N_\eta(\omega-\omega')}+\frac{1}{N_\eta(\omega')N_\xi(\omega-\omega')}\right]f(\omega') \frac{\omega-\omega'}{\omega'}(1-f(\omega-\omega')) + \frac{2f(\omega')(1-f(\omega'))}{N_\xi(\omega')N_\eta(\omega')} \right\}.
\end{align}
At $\kappa = 1$ (or $\kappa=-1$), one can show that $\mu(\omega \ll \Lambda)=0$. Since $\mu=0$ is an unstable fixed point (in the renormalization group sense), the crossover to the stable fixed point can be extracted by considering $\kappa =1 - \delta$ with $\delta$ small. It turns out a scaling form is recovered for $\delta=\sqrt{\pi T^\star/(2\Lambda)}$. In all the frequency integrals in Eq.~\eqref{eq:mu_full_expr}, the contributions to a nonzero $\mu$ are at $\omega \sim T^\star$, so we concentrate on this region and expand in $\omega,T^\star \ll \Lambda$ as
\begin{align}
	f(\omega)&=\frac{\pi|\omega|}{2\Lambda}, \quad N_\eta(\omega)=\frac{\pi(|\omega|+T^\star)}{2\Lambda}, \quad N_\xi(\omega)=4.\label{eq:mu_expansions}
\end{align}
Keeping only the leading terms, the first integral in Eq.~\eqref{eq:mu_full_expr} vanishes and we are left with
\begin{align}
	\mu(i\omega)=&1 -\frac{1}{2 |\omega|}\int d\omega'\left\{\left[\frac{1}{|\omega - \omega'|+T^\star}+\frac{1}{|\omega'|+T^\star}\right] {\rm sgn}(\omega')(\omega-\omega')+\frac{2|\omega'|}{|\omega'|+T^\star}\right\}\nonumber\\
	=&  \frac{T^\star}{|\omega|}\ln\left(1+\frac{|\omega|}{T^\star}\right),
\end{align}
which is $1-\mathcal G^{\rm EK}_{2}(\omega/T^\star )$ with $\mathcal G^{\rm EK}_2$ given in Eq.~(9) of the main text.

\section{\label{app:coulomb_gas}IV. Coulomb gas expansions}
In this section, we review the Coulomb-gas representations of several models belonging either to the Schmid or to the BKT universality class. We emphasize two ingredients that play a crucial role in determining the resulting transition: the dimensionality of the space in which the Coulomb charges reside and the constraints imposed on the charge sequences. The latter point is particularly important in the context of the Schmid transition. Indeed, the distinction between unrestricted and alternating charge sequences was overlooked in Schmid's pioneering analysis~\cite{Schmid_1983_sm} and was subsequently clarified by Bulgadaev~\cite{Bulgadaev_1984_sm}.

The partition function of the boundary sine-Gordon model can be written as a one-dimensional Coulomb gas,
\begin{align}\label{eq:Coulomb_schmid}
	Z=&\sum_{n=0}^\infty y^n\int_{\tau_1<\dots<\tau_n}\sum_{\{s_p=\pm 1\}}\delta\left(\sum_{p=1}^n s_p\right) \exp\left[\gamma\sum_{p<q}s_ps_q\log\left(\frac{\beta}{\pi\tau_c}\sin\left(\pi(\tau_q-\tau_p)/\beta\right)\right)\right]\nonumber.
\end{align}
This representation can be obtained either by expanding in powers of the periodic potential, with $y=\tilde r/2$ and $\gamma=2/\alpha$, or, in the strong-coupling regime, by expanding in instantons, with $y= e^{-S_{\rm inst}}$ and $\gamma=2\alpha$, where $S_{\rm inst}$ is the instanton action. The existence of these two Coulomb-gas representations is a direct manifestation of the infrared self-duality of the bsG model. A crucial feature of Eq.~\eqref{eq:Coulomb_schmid} is that the Coulomb charges reside in a one-dimensional space and their positions are unrestricted: the charges $s_p$ are not constrained to alternate along the sequence. In particular, configurations containing two or more consecutive charges of the same sign are allowed. This property distinguishes the bsG Coulomb gas from the Coulomb gases arising in several other impurity models and is essential for its Schmid rather than BKT critical behavior.

As a contrasting example, consider the one-dimensional Ising model with long-range interactions,
\begin{align}
	H=-\frac{1}{4\alpha}\sum_{i\ne j}\frac{\sigma_i\sigma_j}{\left(\frac{L}{\pi}\right)^2\sin(\pi|i-j|/L)^2}-J\sum_{i=1}^L\sigma_i\sigma_{i+1},
\end{align}
which is equivalent to the single-channel Kondo model~\cite{anderson1969_sm}. Its domain-wall expansion has the same Coulomb-gas form as Eq.~\eqref{eq:Coulomb_schmid}, with $\gamma=2/\alpha$, $y=e^{-J}$, $\tau_c\to1$, and $\beta\to L$, but with a crucial additional constraint: charges must alternate as $s_p=(-1)^p$. This apparently minor modification has profound consequences for the RG flow and the universality class: the model exhibits a BKT transition~\cite{anderson1970_sm,Luijten_2001_sm}.

The dimensionality of the charge space provides another route to BKT behavior. For example, the partition functions of the two-dimensional XY and sine-Gordon models can be expressed as two-dimensional Coulomb gases,
\begin{align}
	Z=&\sum_{n=0}^\infty \frac{y^n}{n!}\int_{\bd r_1,\dots,\bd r_n}\sum_{\{s_p=\pm1\}}\delta\left(\sum_{p=1}^n s_p\right) \exp\left[\gamma\sum_{1\leq p<q\leq n}s_ps_q\log\left(\frac{|\bd r_p-\bd r_q|}{a}\right)\right],
\end{align}
where $a$ is a short-distance cutoff. Here, the charges now reside in a two-dimensional space, $\boldsymbol r_p\in\mathbb R^2$, and consequently there is no natural ordering of the charges. This Coulomb gas undergoes a BKT transition.

\end{document}